\documentclass[aps,pra,twocolumn,groupedaddress,showpacs]{revtex4}

\usepackage{graphicx}
\usepackage{dcolumn}
\usepackage{bm}
\usepackage{verbatim}
\usepackage{amsmath,amssymb}
\usepackage{relsize,exscale}
\usepackage{color}
\begin{document}

\title{Statistical ensembles in Hamiltonian formulation of hybrid quantum-classical systems }

\author{N. Buri\'c}
\email[]{buric@ipb.ac.rs}
\author{I. Menda\v s}
\author{D. B. Popovi\' c}
\author{M. Radonji\'c}
\author{S. Prvanovi\'c}
\affiliation{Institute of Physics, University of Belgrade,
Pregrevica 118, 11080 Belgrade, Serbia}

\begin{abstract}
General statistical ensembles in the Hamiltonian formulation of hybrid
quantum-classical systems are analyzed. It is argued that arbitrary probability
densities on the hybrid phase space must be considered as the class of possible
physically distinguishable statistical ensembles of hybrid systems. Nevertheless,
statistical operators associated with the hybrid system and with the quantum
subsystem can be consistently defined. Dynamical equations for the statistical
operators representing the mixed states of the hybrid system and its quantum
subsystem are derived and analyzed. In particular, these equations irreducibly
depend on the total probability density on the hybrid phase space.
\end{abstract}

\maketitle

Hybrid quantum-classical (QC) systems are neither quantum nor classical. There
is no unique generally accepted theory of the QC systems,
primarily because each of the suggested theories has some unexpected or
controversial features \cite{Else,Kinez,HQC,Anderson,Aleksandrov}.
Current technologies are sufficiently developed to enable
experimental studies of the interaction between typically quantum and typically
classical objects \cite{Exper}. The choice of the proper hybrid theory will
ultimately depend on the experimental tests, but such experiments require
detailed preliminary theoretical models. It is plausible to expect that the
interaction between the classical and the quantum subsystems might introduce
features that are not present neither in the quantum nor in the classical
subsystems without interaction (see for example \cite{Else}). In this
communication we shall explore some features of the hybrid system theory which is formulated using
 the framework of Hamiltonian dynamical systems \cite{Else,usPRA3,Kinez}.

The Hamiltonian hybrid theory, as formulated for example in \cite{Else}, has
many of the properties commonly expected of a good hybrid theory. However, it
also has some controversial features concerning the class of mathematical
objects that should be interpreted as physical variables of the QC system. Also,
what should be mathematical objects that represent the most general statistical
ensembles of QC systems and the corresponding mixtures of the quantum subsystems
is not trivially obvious and needs a careful discussion. Formally, the problem
is inherited from the surplus structure present in the Hamiltonian formulation
of the quantum mechanics. It can be argued that what must be considered as
nonphysical in the Hamiltonian formulation of the standard linear quantum
mechanics might acquire physical meaning for the QC system because of the
presence of the classical subsystems and the quantum-classical interaction.

In what follows we shall first briefly recapitulate the Hamiltonian formulation
of quantum mechanics and of the hybrid quantum-classical systems. Then we shall
discuss possible representation of general statistical ensembles of QC systems
within the Hamiltonian formulation. Statistical operator for the QC system, as
well as conditional and unconditional mixed  states of the quantum subsystem
corresponding to the general ensembles of the QC systems, will be defined and
their evolution will be discussed.

{\it Hamiltonian formulation of quantum mechanics}

Schr\"odinger dynamical equation on a Hilbert space ${\cal H}$ generates a
Hamiltonian dynamical system on an appropriate symplectic manifold
\cite{HamiltonianQM,us}. The real manifold ${\cal M}$, associated with the Hilbert
space ${\cal H}$ in fact has Riemannian and symplectic structure, provided by
the real and the imaginary parts of the scalar product, and can be viewed as a
phase space of a Hamiltonian dynamical system, additionally equipped with the
Riemannian metric which reflects its quantum origin. A vector $|\psi\rangle$ from ${\cal H}$,
associated with a pure quantum state, is represented by the corresponding point
in the phase space ${\cal M}$ denoted by $X_{\psi}$ or simply by $X$.

Real coordinates $\{(x_j,y_j)$, $j=1,2,\dots \}$ of a point $\psi\in{\cal H}
\equiv{\cal M}$ are introduced using expansion coefficients $\{c_j$,
$j=1,2,\dots\}$ in some basis $\{|j\rangle,$ $j=1,2,\dots\}$ of ${\cal H}$ as
follows\vspace{-1mm}
\begin{equation}
|\psi\rangle=\sum_j c_j|j\rangle,\: x_j=\sqrt{2}\,{\rm Re}\,c_j,\:y_j=\sqrt{2}\,{\rm Im}\,c_j.
\end{equation}\vspace{-4mm}\\
The coordinates $(x_j,y_j)$ represent canonical coordinates of a Hamiltonian
dynamical system on ${\cal M}$. Consequently, the Poisson bracket between two
functions $F_1$ and $F_2$ on ${\cal M}$ in the canonical coordinates $(x_j,y_j)$
is given by\vspace{-1mm}
\begin{equation}
\{F_1,F_2\}_{\cal M}=\frac{1}{\hbar}\sum_j \left(\frac{\partial F_1}{\partial x_j}\frac{\partial
F_2}{\partial y_j}-\frac{\partial F_2}{\partial x_j}\frac{\partial F_1}{\partial y_j}\right).
\end{equation}\vspace{-3mm}\\
A quantum observable $\hat H$ is represented by the corresponding function
of the form\vspace{-1mm}
\begin{equation}
H(X_{\psi})=\langle\psi|\hat H|\psi\rangle.
\end{equation}\vspace{-5mm}\\
Hamiltonian flows with the
Hamilton's function of the form (3) generate isometries of
the Riemannian metric. More general Hamiltonian flows on
${\cal M}$, corresponding to the Hamilton's function which are not
of the form (3), do not generate isometries and do not have the physical
interpretation of quantum observables. It can be seen easily that\vspace{-2mm}
\begin{equation}
\{H_1,H_2\}_{\cal M}=\frac{1}{i\hbar}\langle[\hat H_1,\hat H_2]\rangle.
\end{equation}\vspace{-4mm}\\
The Schr\"odinger evolution equation\vspace{-1mm}
\begin{equation}
i\hbar|\dot\psi\rangle=\hat H|\psi\rangle
\end{equation}\vspace{-5mm}\\
is equivalent to the Hamilton equations on ${\cal M}$  assuming the standard form
in the canonical coordinates $(x_j,y_j)$\vspace{-2mm}
\begin{equation}
\dot x_j=\frac{\partial H}{\partial y_j},\qquad \dot y_j=-\frac{\partial H}{\partial x_j},
\end{equation}\vspace{-3mm}\\
with $H$ given by (3).

The Hamiltonian formulation of the Schr\"odinger equation on ${\cal H}$
automatically preserves the constraints imposed by the physical equivalence of
Hilbert space vectors. In fact, Hamiltonian formulations based on ${\cal H}$ and
on the projective Hilbert space as the space of physical states are equivalent.
We use the formulation in which points of the quantum phase space are identified
with the vectors from ${\cal H}$ since it is sufficient for our main purpose.

{\it Mixed states of a quantum system in the Hamiltonian formulation}

A quantum state is in general represented by the corresponding density operator
$\hat\rho$ on ${\cal H}$. On the other hand every positive function $\rho(x,y)$
with unit integral on ${\cal M}$ represents a density of some probability theory on
${\cal M}$. Expectation of a function $F(x,y)$ with respect to $\rho(x,y)$ is
given by\vspace{-2mm}
\begin{equation}
\bar F=\int_{\cal M} \rho(x,y) F(x,y)dM,
\end{equation}\vspace{-3mm}\\
where $dM$ is the Lebesque  measure on ${\cal M}$.
The densities satisfy Liouville equation on ${\cal M}$\vspace{-2mm}
\begin{equation}
\frac{\partial}{\partial t}\rho(x,y;t)=\big\{H(x,y),\rho(x,y;t)\big\}_{\cal M}.
\end{equation}\vspace{-4mm}\\
Quantum mechanical average of the observable $\hat F$ in the state $\hat\rho$,
${\rm Tr}(\hat\rho\hat F)$, is reproduced with the formula (7) using any of the
probability densities $\rho(x,y;t)$ with the same first moment fixed by the
requirement\vspace{-1mm}
\begin{equation}
\hat\rho(t_0)=\int_{\cal M}\rho(x,y;t_0)\hat\Pi(x,y)dM,
\end{equation}\vspace{-3mm}\\
where $\hat\Pi(x,y)=|\psi_{x,y}\rangle\langle\psi_{x,y}|$ and the state $|\psi_{x,y}
\rangle\in{\cal H}$ corresponds to the coordinates $(x,y)$ of ${\cal M}$. Liouville
evolution of the densities $\rho(x,y;t)$ yielding the same $\hat\rho(t_0)$ generates
the same von Neumann evolution $\hat\rho(t)$. The fact that the quantum mixed state
$\hat\rho$ determines only an equivalence class of densities, those with the appropriate
first moment, is equivalent to the non-uniqueness of the expansion of the quantum
mixed state in terms of convex combinations of pure state projectors.

{\it Hamiltonian theory of hybrid systems}

Hamiltonian theory of hybrid quantum-classical systems can be developed starting
from the Hamiltonian
formulation of a composite quantum system and imposing a constraint that one of
the components is behaving as a
classical system \cite{usPRA3}. The result in the macro-limit imposed on the
classical subsystem turns out to be
 equivalent to a Cartesian product of two Hamiltonian systems as in
\cite{Else}. One of these Hamiltonian systems corresponds to
 the quantum and one to the classical subsystems of the hybrid. However, the
interaction between the two subsystems
has crucial influence on their properties.

The phase space of the hybrid system ${\cal M}$ is considered as a Cartesian
product ${\cal M}={\cal M}_c\times {\cal M}_q$ of the classical subsystem phase
space ${\cal M}_c$ and of the quantum subsystem phase space ${\cal M}_q$.
Denoting the local coordinates on the product as $\{p,q,x,y\}$,  where $(p,q)\in
{\cal M}_c$ and $(x,y)\in {\cal M}_q$ one can write the evolution equations of
the QC system as Hamiltonian dynamical equations on the phase space $\cal M$
with the Hamilton's function comprised of three terms
\begin{equation}
H_t(p,q,x,y)=H_c(p,q)+H_q(x,y)+V_{int}(p,q,x,y),
\end{equation}
where $H_c$ is the Hamilton's function of the classical subsystem, $H_q(x,y)$ of
the form (3) is the Hamilton's function of the
quantum subsystem and $V_{int}(p,q,x,y)=\langle \psi_{x,y}|\hat
V_{int}(p,q)|\psi_{x,y}\rangle$, where $\hat V_{int}(p,q)$
is an operator in the Hilbert space of the quantum subsystem which depends on
the classical coordinates $(p,q)$ and describes the
interaction between the subsystems. The Poisson bracket on $\cal M$ of arbitrary
functions of the local coordinates $(p,q,x,y)$ is defined as\vspace{-2mm}
\begin{align}
\{f_1,f_2\}_{\cal M}&=\sum_{i=1}^k\left(\frac{\partial f_1}{\partial q_i}
\frac{\partial f_2}{\partial p_i}-\frac{\partial f_2}{\partial q_i}
\frac{\partial f_1}{\partial p_i}\right )\nonumber\\
&+\frac{1}{\hbar}\sum_j\left(\frac{\partial f_1}{\partial x_j}
\frac{\partial f_2}{\partial y_j}-\frac{\partial f_2}{\partial x_j}
\frac{\partial f_1}{ \partial y_j}\right)\!.
\end{align}\vspace{-3mm}\\
Thus, the Hamiltonian form of the QC dynamics on
$\cal M$ as the phase space reeds
\begin{eqnarray}
&&\dot q=\{q,H_t\}_{\cal M},\quad \dot p=\{p,H_t\}_{\cal M},\\
&&\dot x=\{x,H_t\}_{\cal M},\quad \dot y
=\{y,H_t\}_{\cal M},
\end{eqnarray}
where the Hamilton's function $H_t(p,q,x,y)$ in local
coordinates on $\cal M$ is given by (10).

{\it Statistical ensembles of QC systems and quantum subsystems}

Consider a general probability density $\rho(p,q,x,y)$ on the total hybrid phase
space ${\cal M}={\cal M}_c\times {\cal M}_q$. There is no reason to require such
probability density to represent a physical quantity or an observable of the QC system.
If such $\rho(p,q,x,y)$ is a quadratic function of $x,y$ then it is equal to
the quantum expectation in the corresponding state $|\psi_{x,y}\rangle$ of an
operator function $\hat f_{\rho}(p,q)$, i.e., $\rho(p,q,x,y)=
\langle\psi_{x,y}|\hat f_{\rho}(p,q)|\psi_{x,y}\rangle$, where for each fixed
$p,q$ the operator $\hat f_{\rho}(p,q)$ is a statistical operator on the Hilbert
space of the quantum subsystem. However, in general a probability density of an
arbitrary form $\rho(p,q,x,y)$ describes a perfectly legitimate statistical
ensemble of QC systems. In general, following the  Hamiltonian formulation of
the QC system dynamics, the evolution of $\rho(p,q,x,y;t)$ considered as a
statistical ensemble on ${\cal M}$ is given by the Liouville equation with the
Hamilton's function (10) and the Poisson bracket (11)\vspace{-1mm}
\begin{equation}
\frac{\partial}{\partial t}\rho(p,q,x,y;t)=\big\{H_t(p,q,x,y),\rho(p,q,x,y;t)\big\}_{\cal M},
\end{equation}\vspace{-3mm}\\
i.e., $\rho(p(t),q(t),x(t),y(t);t)\!=\!const$ when $(p(t),q(t))$ and $(x(t),y(t))$
are determined from the Hamilton equations (12) and (13), respectively.
However, Liouville evolution of an ensemble which is at $t=t_0$ of the form
$\langle\psi_{x,y}|\hat f_{\rho}(p,q;t_0)|\psi_{x,y}\rangle$ will in general
result in some probability density $\rho(p,q,x,y;t)$ which is not quadratic in
$x,y$, i.e., can not be expressed as expectation of an operator. Therefore,
it can be argued that the most general statistical ensembles of QC systems
need to be represented by general probability densities $\rho(p,q,x,y;t)$. We
shall therefore assume, in accordance with the Hamiltonian theory, that an
arbitrary probability density $\rho(p,q,x,y;t)$ describes a statistical ensemble
of QC systems, that is a mixed state of the hybrid, and that the evolution of
such mixed states is given by the corresponding Liouville equation.

Suppose a QC system is in a general mixed state $\rho(p,q,x,y;t)$.
The density $\rho(p,q,x,y;t)$ generates a unique positive operator valued
function (POVF):\vspace{-1mm}
\begin{equation}
\hat\rho(p,q;t)=\int_{{\cal M}_q}\rho(p,q,x,y;t)\hat\Pi(x,y)dM_q,
\end{equation}\vspace{-2mm}\\
which can be called the hybrid statistical operator. $\hat\rho(p,q;t)$ contains
less information about the hybrid system state then the density $\rho(p,q,x,y;t)$,
and plays a secondary role in the hybrid theory presented here. The corresponding
mixed state of the quantum subsystem conditional on the classical subsystem being
in the state $(p,q)$ is uniquely represented by\vspace{-2mm}
\begin{equation}
\hat\rho_{p,q}(t)=\hat\rho(p,q;t)/\int_{{\cal M}_q}\rho(p,q,x,y;t)dM_q.
\end{equation}\vspace{-4mm}\\
The unconditional mixed state of the quantum subsystem of the hybrid in the
state $\rho(p,q,x,y;t)$ is also uniquely obtained as\vspace{-1mm}
\begin{equation}
\hat\rho(t)=\int_{\cal M} \rho(p,q,x,y;t)\hat\Pi(x,y) dM.
\end{equation}\vspace{-4mm}\\
At time $t$ the previous formula defines positive, trace one operator, i.e.,
a statistical operator representing the mixed state of the quantum subsystem.
At any $t$ the statistical operator $\hat\rho(t)$ depends on the value of the
sub-integral expression at the same time $t$. The Liouville evolution of
$\rho(p,q,x,y;t)$ is certainly continuous in $t$ and the dependence on $t$ of
$\hat\rho(t)$ given by (17) is also continuous. Thus, the formula (17) defines a
continuous one-parameter family of statistical operators on ${\cal H}$.

Analogously to the relation (9) valid for a purely quantum system, many hybrid
ensembles represented by different $\rho(p,q,x,y;t)$ have the
quantum subsystem in the same conditional or unconditional mixed state. The
crucial difference between the purely quantum and the hybrid
systems is that we have assumed that each different $\rho(p,q,x,y;t)$ describes
physically different ensembles of QC systems with the quantum subsystem in the
same mixed state. This will be reflected in the evolution of (15) or (17).
Recall that all different $\rho(x,y;t)$ in (9) with the same first
moment correspond to the physically equivalent quantum mixture $\hat\rho(t_0)$,
and generate unique von Neumann evolution of $\hat\rho(t)$ which is obtained
from the Liouville evolution of any such $\rho(x,y;t)$.
Therefore, all such $\rho(x,y;t)$ are equivalent in the purely quantum case. In
the hybrid case, different $\rho(p,q,x,y;t)$ which give the same $\hat\rho(p,q;t_0)$
(or $\hat\rho(t_0)$), as we shall see, generate different evolution of
$\hat\rho(p,q;t)$ (or $\hat\rho(t)$) and thus must be considered as physically
different.

The evolution equation satisfied by $\hat\rho(p,q;t)$ can be obtained from (14)
and (15) using partial integration over $(x_j,y_j)$ and the identities $\partial F/
\partial x_j=(\langle\psi_{x,y}|\hat{F}|j\rangle+\langle\;\!\!j|\hat{F}|\psi_{x,y}\rangle)
/\sqrt{2}$, $\partial F/\partial y_j=i(\langle\psi_{x,y}|\hat{F}|j\rangle-\langle\,j|
\hat{F}|\psi_{x,y}\rangle)/\sqrt{2}$, where $F=\langle\psi_{x,y}|\hat{F}|\psi_{x,y}\rangle$.
The resulting equation is
\begin{widetext}\vspace*{-6mm}
\begin{align}
\frac{\partial\hat\rho(p,q;t)}{\partial t}=
\frac{1}{i\hbar}\big[\hat H_q\!+\!\hat V_{int}(p,q),\hat\rho(p,q;t)\big]
\!+\!\big\{H_c(p,q),\hat\rho(p,q;t)\big\}_{p,q}
\!+\!\int_{{\cal M}_q}\!\!\big\{V_{int}(p,q,x,y),\rho(p,q,x,y;t)\big\}_{p,q}\,\hat\Pi(x,y)dM_q.
\end{align}
\end{widetext}
The solution of (18) remains a well defined statistical operator on ${\cal H}$
for all $t$, which is a desirable property not shared by some other hybrid system
theories \cite{Anderson, Aleksandrov}. The equation for the statistical operator of
the quantum subsystem $\hat\rho(t)$ follows after the integration over $(p,q)$
\begin{widetext}\vspace{-4mm}
\begin{align}
\frac{d\hat\rho(t)}{dt}&=\frac{1}{i\hbar}\big[\hat H_q,\hat\rho(t)\big]+
\frac{1}{i\hbar}\int_{{\cal M}_c}\!\!\big[\hat V_{int}(p,q),\hat\rho(p,q;t)\big]dM_c
+\int_{{\cal M}_c}\big\{H_c(p,q),\hat\rho(p,q;t)\big\}_{p,q}dM_c\nonumber\\
&+\int_{{\cal M}}\big\{V_{int}(p,q,x,y),\rho(p,q,x,y;t)\big\}_{p,q}\,\hat\Pi(x,y)dM.
\end{align}
\end{widetext}\vspace*{-3mm}
The first  term on the right side of (19) generate the unitary part of the
evolution. The last three terms do not preserve the norm of $\hat\rho$, and are responsible for non-unitary  effects.
Notice that the evolution of $\hat\rho(p,q;t)$ ($\hat\rho(t)$) cannot be expressed
only in terms of $\hat\rho(p,q;t)$ ($\hat\rho(t)$), but irreducibly involves the
probability density $\rho(p,q,x,y;t)$.

Observe that, taking different $\rho'(p,q,x,y;t)$ yielding the same
$\hat\rho'(p,q;t_0)=\hat\rho(p,q;t_0)$ via (15) will in general generate
different $\hat\rho'(p,q;t) \neq\hat\rho(p,q;t)$. In other words, the states of
the quantum subsystem of a hybrid in different states $\rho(p,q,x,y;t)$ and
$\rho'(p,q,x,y;t)$ might be the same at some moment $t_0$, but will inevitably
evolve differently. This is natural since, the corresponding evolution equation
(18) for $\hat\rho(p,q;t)$ must depend on the evolution of the entire QC system.
In particular, one might adjust the total initial ensemble of the hybrid so that the
evolution of the quantum subsystem from a fixed initial mixture $\hat\rho(t_0)$ has
different properties, without altering the Hamiltonian. Experimental observation of
different evolutions of the same initial quantum state $\hat\rho(p,q;t_0)$, obtained
from multiple $\rho(p,q,x,y;t)$ that are different functions of $(x,y)$, would
provide a confirmation of our main assumption concerning the class of physically
distinguishable ensembles of hybrid QC systems.

{\it Summary and discussion}

In summary, we have explored some of the consequences of the assumption that in
the Hamiltonian formulation the set of ensembles of hybrid quantum-classical systems
is mathematically represented by the space of probability
densities on the hybrid system phase space. Each such ensemble uniquely
determines a conditional and an unconditional mixed state of the quantum
subsystem represented by the corresponding density operators on the quantum
subsystem Hilbert space. Evolution of the quantum subsystem mixtures is defined
using the evolution of the hybrid probability density. Different hybrid ensembles
might give the same quantum subsystem mixture at some time $t_0$, but that
quantum mixture obtained from different hybrid ensembles will evolve differently.
The evolution equations for the hybrid statistical operator $\hat\rho(p,q;t)$ (18)
and for the quantum subsystem unconditional mixture $\hat\rho(t)$ (19)
have been derived and inevitably involve the full density $\rho(p,q,x,y;t)$.

The initial assumption about the mathematical objects needed to represent all
physically possible ensembles of hybrid systems in fact assumes
that the ensembles of possibly interacting quantum-classical systems are more
general then ensembles of quantum-classical systems without the
 interaction between the subsystems. In the later situation a set of densities
on the hybrid phase space that is invariant under the evolution can be chosen to
contain only function that are necessary of quadratic dependence on the quantum
degrees of freedom. However, if the quantum and the classical subsystem interact
than the invariant set of densities is in general the full set of probability
densities on the hybrid phase space. We have shown how such general set of hybrid
ensembles generates consistently defined quantum mechanical mixtures of the
quantum subsystem with the corresponding evolution equations reflecting the
quantum-classical interaction. The interaction implies that the evolution of the
quantum subsystem statistical operator shows explicit dependence on the
equivalent representations of the initial density operator.

{\bf Added note} Formula (19) can be considerably simplified since the last two terms containing integrals over classical
 variables of the Poison brackets are in fact equal to zero. However, this does not effect the main
  conclusions that the evolution of $\hat\rho(t)$ depends on the full hybrid density and is non-unitary, since these conclusions
   are consequences of the term with the commutator of the interaction $\hat V(p,q)$ and $\hat\rho(p,q)$. 
   Only if there is no interaction between the classical and the quantum degrees of freedom the evolution of the quantum density $\hat\rho(t)$
    is unitary and independent of its initial expression in terms of the full hybrid density.  

\acknowledgments
This work was supported by the Ministry of Science and Education of the
Republic of Serbia, contracts No.\ 171006, 171017, 171020, 171028 and 171038
and by COST (Action MP1006).

\end{document}